\documentclass[]{spie}  
\usepackage[]{graphicx}

\usepackage{amsmath,amssymb,amsfonts,mathrsfs} 
\usepackage{subfigure}
\usepackage{fancyhdr} 
\usepackage{lastpage} 
\usepackage{url} 
\usepackage{array}
\usepackage{tabularx}
\usepackage{supertabular}
\usepackage{multirow}
\usepackage{rotating}
\usepackage{nomencl}
\usepackage{bm}

\usepackage{color} 
\usepackage{epsfig}

\title{Extremely fast focal-plane wavefront sensing for extreme adaptive optics}

\author{Christoph U. Keller\supit{a}, Visa Korkiakoski\supit{a}, Niek Doelman\supit{b}, Rufus Fraanje\supit{c}, Raluca Andrei\supit{c} and Michel Verhaegen\supit{c}
\skiplinehalf
\supit{a}Leiden Observatory, Niels Bohrweg 2, 2333CA Leiden, The Netherlands; \\
\supit{b}TNO Science and Industry, Stieltjesweg 1, 2628CK Delft, The Netherlands; \\
\supit{c}Delft Center for Systems and Control, Mekelweg 2, 2628CD Delft, The Netherlands \\
}

\authorinfo{E-mail: keller@strw.leidenuniv.nl}

\begin{document} 
\maketitle 

\begin{abstract}
  We present a promising approach to the extremely fast sensing and
  correction of small wavefront errors in adaptive optics systems. As
  our algorithm's computational complexity is roughly proportional to
  the number of actuators, it is particularly suitable to systems with
  10,000 to 100,000 actuators. Our approach is based on sequential
  phase diversity and simple relations between the point-spread
  function and the wavefront error in the case of small
  aberrations. The particular choice of phase diversity, introduced by
  the deformable mirror itself, minimizes the wavefront error as well
  as the computational complexity. The method is well suited for
  high-contrast astronomical imaging of point sources such as the
  direct detection and characterization of exoplanets around stars,
  and it works even in the presence of a coronagraph that suppresses
  the diffraction pattern. The accompanying paper in these proceedings
  by Korkiakoski et al. describes the performance of the algorithm
  using numerical simulations and laboratory tests.
\end{abstract}

\keywords{Adaptive optics, wavefront sensing, sequential phase diversity}

\section{Introduction}
\label{sec:intro} 

The computational requirements for extreme adaptive optics for the
next generation of 30 to 40-meter class astronomical telescopes
present a formidable problem. For an adaptive optics system with a
phase corrector that has $N$ degrees of freedom, classical approaches
require a multiplication of a vector of length $N$ with a matrix of
size $N\times N$, leading to a computational complexity of order
$N^2$. For extreme adaptive optics, this vector-matrix multiplication
is the major bottleneck, not only in terms of computational power but
also in terms of memory requirements as both scale with $N^2$. Indeed,
a system with $N=40,000$ as envisaged for the EPICS instrument at the
E-ELT, the classical reconstruction matrix would require about 6GB of
memory (assuming 4-byte single-precision floating point numbers). This
data needs to be moved into the processing units at a rate of a few
thousand times per second, making this issue alone a challenging data
flow problem.

Even when using parallel processing and distributed memory techniques,
the most modern hardware implementation would struggle to achieve the
computational complexity and speed. Depending on the type of wavefront
sensor and deformable mirror, sparse-matrix
approaches\cite{gilles2002}, sequential one-dimensional
reconstructions\cite{obereder2011} and Fourier reconstruction
techniques\cite{poyneer2002} can reduce the required processing
power. Typically, these approaches work with conventional pupil-based
wavefront sensors such as Shack-Hartmann and pyramid wavefront
sensors.

Here we present a focal-plane sensing algorithm with a complexity that
is proportional to $N\log N$ in terms of required computing power and
proportional to $N$ in terms of memory requirements. It is based on 1)
the close relationship between the PSF and the wavefront aberration in
case that the latter is small\cite{ellerbroek1983, gonsalves2001,
  sivaramakrishnan2002} , 2) sequential phase-diversity
\cite{gonsalves2002, gonsalves2010} where the deformable mirror itself
introduces the phase diversity and 3) a choice of the introduced phase
diversity that minimizes the wavefront error and fortunately also
minimizes the computational effort. Indeed, our algorithm only
requires a few floating point calculations per wavefront resolution
element and a single, two-dimensional, complex Fourier transform per
update cycle.  Our work combines knowledge from the
areas of AO-corrected speckle studies\cite{sivaramakrishnan2002} and
the phase-diversity efforts of
Gonsalves\cite{gonsalves2001,gonsalves2002}.

We begin in Sect.~\ref{sec:psf} by calculating the monochromatic PSF
of a weakly aberrated system to second order, which allows us to
derive a correction to the first-order approximation that makes it
significantly better.  We continue in Sect.~\ref{sec:spd} by deriving
the basic sequential phase diversity technique for our
approximation. Sect.~\ref{sec:complex} shows how the computational
complexity can be dramatically reduced from four Fourier transforms to
a single one. The advantages of our approach are discussed in
Sect.~\ref{sec:adv}, the limitations are listed in
Sect.~\ref{sec:lim}, and applications and extensions are summarized in
Sect.~\ref{sec:apps}.  In a separate paper in these proceedings, we
have also studied in detail one version of our algorithm numerically
as well as in the laboratory with a 37-actuator system
\cite{korkiakoski2012spie}.

\section{Monochromatic PSF due to weak aberrations}
\label{sec:psf}

The overall approach is based on the work by
Gonsalves\cite{gonsalves2001}, who presented an analytical solution to
retrieving the wavefront from two simultaneous images of a point
source where one is in focus and the other one has a phase diversity,
typically defocus, applied. Gonsalves\cite{gonsalves2002,
  gonsalves2010} also introduced sequential phase diversity (SPD)
where the phase-diverse images are acquired sequentially instead of
simultaneously. We expand Gonsalves' work here to second order with a
notation that is more closely related to the work that has been done
for understanding the PSF of AO-corrected point
sources\cite{perrin2003}. This expansion leads to a much deeper
understanding of the approach suggested by
Gonsalves\cite{gonsalves2001} and implies a modification that
significantly expands the range of the weak aberration approximation
as shown in the accompanying paper by
Korkiakoski\cite{korkiakoski2012spie}.

\subsection{First-order approximation of wavefront}

We start with the (scalar) electrical field of an electromagnetic wave in the
pupil plane of an optical system,
\begin{equation}
E(u,v)=A(u,v) e^{i\Phi(u,v)}
\label{eq:pupil}
\end{equation}
where $A(u,v)$ is the amplitude and $\Phi(u,v)$ is the phase in radians, both
being functions of the spatial coordinates $u$ and $v$ in the pupil
plane. The origin of the $(u,v)$ coordinate system is the center of
the aperture. Both $A(u,v)$ and $\Phi(u,v)$ are real. Furthermore, as
the absolute value of the phase (piston term) is not relevant, we
can set the average of $\Phi$ over the aperture to zero, i.e.\
\begin{equation}
\int A(u,v)\Phi(u,v)\; du\; dv = 0\,.
\label{eq:norm}
\end{equation}
Note that $A(u,v)$ does not have to be limited to values of 0 or 1 but
can have values in between such as used in apodized-pupil
coronagraphs\cite{kasdin2003}.

As we do not rely on absolute intensity measurements and assume that
we only deal with pure phase aberrations, the amplitude of
the electrical field is normalized such that
\begin{equation}
\int |E(u,v)|^2du\;dv = \int |A(u,v)|^2\; du\;dv = 1\,.
\end{equation}
Furthermore, we assume that the amplitude is even, i.e.\
\begin{equation}
A(u,v)=A(-u,-v)\;.
\end{equation}
This is an assumption that will greatly simplify the subsequent
calculations without being overly restrictive. An even amplitude or
aperture function $A(u,v)$ still describes most
telescope apertures but also holds for shaped pupils and
apodized pupils that are used to suppress diffraction
effects around a point source\cite{kasdin2003}.

If the aberrations are very small, i.e.\ $\Phi\ll 1$, $E(u,v)$ in Eq.(\ref{eq:pupil})
can be approximated to first order by 
\begin{equation}
E(u,v)\approx E^\prime(u,v) = A(u,v)\left[1+i\Phi(u,v)\right]\,.
\label{eq:smallapprox}
\end{equation}

We note that the integral over the absolute value squared of
$E^\prime$ is not unity anymore since
\begin{equation}
\int |E^\prime|^2\; du\;dv = \int A(u,v)^2 \left[1+\Phi(u,v)^2\right] du\;dv =
1+\int \left[A(u,v) \Phi(u,v)\right] ^2\; du\;dv\,.
\end{equation}
The error made in the normalization of the amplitude by this
first-order approximation is therefore given by the variance of the
aperture-weighted wavefront, $A\Phi$ because of
Eq.(\ref{eq:norm}). This violation of energy conservation is the
fundamental reason that algorithms based on
Gonsalves\cite{gonsalves2001} first-order approximation are limited to
extremely small phase aberrations and that the Strehl ratio of the
corresponding PSF is always 1, independent of the wavefront variance,
as will be shown below.

\subsection{Second-order approximation of wavefront}

Expanding Eq.(\ref{eq:pupil}) to second order, we obtain 
\begin{equation}
E(u,v)\approx E^{\prime\prime}(u,v) =
A(u,v)\left[1+i\Phi(u,v)-\frac{1}{2}\Phi(u,v)^2\right]\;.
\label{eq:e2nd}
\end{equation}
The integral
\begin{equation}
\int |E^{\prime\prime}(u,v)|^2\;du\;dv = \int A(u,v)^2 \left[\left(1
-\frac{1}{2}\Phi(u,v)^2\right)^2+\Phi(u,v)^2\right]du\;dv \approx 1
\end{equation}
up to second order for $\Phi\ll 1$ since $(1-\frac{x}{2})^2 \approx 1-x)$ for
$x\ll 1$. Therefore, expanding Eq.(\ref{eq:pupil}) up to second order
in the wavefront aberration $\Phi$ will guarantee energy conservation
up to second order.

In the first-order approximation of the complex electrical field in
the aperture, the real term is simply 1. In the second-order
approximation, the real term of the series becomes
$1-\frac{1}{2}\Phi(u,v)^2$, which will always be smaller than 1.  The
first-order expansion will therefore always overestimate the real
term. We can correct for this overestimation in a statistical sense by
replacing the general $\frac{1}{2}\Phi(u,v)^2$ term in the complex
field amplitude approximation with its average (over the pupil),
\begin{equation}
\frac{1}{2}\Phi(u,v)^2 \approx \frac{1}{2}\int A(u,v)^2\Phi(u,v)^2\; du\;dv = \frac{1}{2}\sigma_\Phi^2\,,
\end{equation}
where $\sigma_\Phi^2$ is the wavefront variance in radians squared. We
can therefore mitigate the problem in the first-order approximation of
the pupil electric field by scaling the real part with
$1-\sigma_\Phi^2/2$. The imaginary part will not require any scaling.

We note that the square of this scaling factor, to second order in $\Phi$
\begin{equation}
(1-\sigma_\Phi^2/2)^2 \approx 1-\sigma_\Phi^2
\end{equation}
is the same as the Strehl ratio for small aberrations, the extended
Marechal approximation\cite{ross2009}. At the end of this subsection,
we will provide an alternative explanation for this scaling factor
that is closely linked to the second-order expansion of the wavefront
aberration and provides a clear connection between this term and the
Strehl ratio.

\subsection{Fourier transforms and symmetries}

Following Gonsalves\cite{gonsalves2001} we split the phase $\Phi$ into odd and
even terms such that
\begin{equation}
\Phi(u,v) = \Phi_o(u,v)+\Phi_e(u,v)
\end{equation}
and
\begin{equation}
\Phi_o(u,v)=-\Phi_o(-u,-v),\ \ \Phi_e(u,v)=\Phi_e(-u,-v)\;.
\end{equation}

The corresponding approximate complex electric field amplitude in the
focal plane is given by the Fourier transform of
$E^{\prime\prime}(u,v)$ in Eq.(\ref{eq:e2nd}). The coordinate system in
the focal plane has its origin at the center of the perfect PSF
without aberrations. We will not explicitly write the dependence on
the focal-plane coordinates to make the equations more
readable. Hence, the electrical field in the focal plane to second
order in the phase aberration $\Phi$ is
\begin{equation}
e^{\prime\prime} = a+ia*\left(\phi_e+\phi_o\right)-\frac{1}{2}a*\left(\phi_e+\phi_o\right)*\left(\phi_e+\phi_o\right)\;,
\end{equation}
where lower-case symbols are the Fourier transforms of the upper-case
symbols, and $*$ is the convolution operator.

The Fourier transform of a real, even function is real and even, and the
Fourier transform of a real, odd function is a purely imaginary, odd
function. Therefore, $\phi_e$ and $a$ are real and even, and $\phi_o$ is
imaginary and odd.

\subsection{Point-Spread Function}

The point-spread function (PSF) to second order in the aberration,
$p$, is given by the absolute value squared of the complex field
amplitude in the focal plane,
\begin{eqnarray}
p &=& e^{\prime\prime}\cdot e^{\prime\prime *} \\
   &=& a^2 \\ \nonumber
   && +ia\left[a*(\phi_e+\phi_o)-a*(\phi_e^*+\phi_o^*)\right]\\ \nonumber
   &&
   +\left[a*(\phi_e+\phi_o)\right]\left[a*(\phi_e^*+\phi_o^*)\right]
   \\ \nonumber
   && -\frac{1}{2}a\left[a*(\phi_e+\phi_o)*(\phi_e+\phi_o)+
                          a*(\phi_e^*+\phi_o^*)*(\phi_e^*+\phi_o^*)\right]
                        \;, \nonumber
\label{eq:psf2nd}
\end{eqnarray}
where $^*$ indicates the complex conjugate and we have dropped terms
proportional to $\Phi^3$ and higher.  With $A(u,v)$ being real and even, $a=a^*$ is real and
$a^2$ is simply the PSF without aberrations. Since
$\phi_e^*=\phi_e$ and $\phi_o^*=-\phi_o$, Eq.(\ref{eq:psf2nd})
simplifies to
\begin{eqnarray}
p(x,y) &=& a^2 \\ \nonumber
   && +2ia(a*\phi_o) \\ \nonumber
  && +(a*\phi_e)^2-(a*\phi_o)^2 \\ \nonumber
  && -a\left(a*\phi_o*\phi_o+a*\phi_e*\phi_e\right)\;. \nonumber
\label{eq:psf3}
\end{eqnarray}
This is the PSF to second order in the aberrations $\Phi$. Note that
the last term is mostly missing in publications that use the first-order
approximation in the wavefront to derive an approximate PSF. 

Using the notation from Gonsalves'\cite{gonsalves2001} work on
phase-diversity in the weak aberration limit,
\begin{eqnarray}
v &=& a*\phi_e\\
y &=& ia*\phi_o
\end{eqnarray}
and both $v$ and $y$ being real quantities, we can rewrite
Eq.\ref{eq:psf3} as
\begin{equation}
p=a^2+2ay+y^2+v^2-a\left(a*\phi_e*\phi_e+a*\phi_o*\phi_o\right)\;.
\end{equation}

Following again Gonsalves\cite{gonsalves2001}, we separate the PSF, $p$, into its odd
and even parts,
\begin{eqnarray}
p_o&=&2ay\\ \label{eq:po}
p_e&=&
a^2+y^2+v^2-a\left(a*\phi_e*\phi_e+a*\phi_o*\phi_o\right)\,. \label{eq:pe}
\end{eqnarray}
The last term for the even part of the wavefront is difficult to
include in analytical solutions of the wavefront sensing problem as it
includes convolutions of wavefronts with themselves and the
aperture. However, it is the only term that can be negative and that
is not zero at the center of the unaberrated PSF, the coordinate
origin in the focal plane. The latter is due to our previous
assumption that the phase, averaged over the aperture, is zero. It is
indeed this term that leads to a reduction of the Strehl ratio as has
already been pointed out by Perrin et al.\cite{perrin2003}.

Due to the requirement for energy conservation, the integral over the
PSF has to remain constant, independent of the wavefront
aberration. Therefore
\begin{equation}
\int
y^2+v^2-a\left(a*\phi_e*\phi_e+a*\phi_o*\phi_o\right)\,dx\;dy=0\,.
\end{equation}
Since the first two terms are always positive, the third term has to
be largely negative. Neglecting this term therefore leads to a
consistent bias in the estimate for the amplitude of $v$, which limits
the applicability of Gonsalves' algorithm to very small phase
aberrations.

To simplify matters, we approximate $\Phi(u,v)^2$ with its average
$\sigma_\Phi^2$. $\phi*\phi$ therefore becomes a delta function and
the last term in Eq.(\ref{eq:pe}) becomes
\begin{equation}
a\left(a*\phi_e*\phi_e+a*\phi_o*\phi_o\right)\approx
\sigma_\Phi^2 a^2\,.
\end{equation}
Finally, the equation for the even part of the PSF reduces to
\begin{equation}
p_e= (1-\sigma_\Phi^2)a^2+y^2+v^2\,.
\label{eq:pesc}
\end{equation}

The even part is a modified version of the equation given by
Gonsalves\cite{gonsalves2001} where we correct the unaberrated PSF,
$a^2$, with the factor $1-\sigma_\Phi^2$. Even though we do not know
the variance of the wavefront aberrations in advance of the actual
wavefront sensing, we can easily determine this correction factor as
it is the same as the Strehl ratio for small aberrations in the
extended Marechal approximation\cite{ross2009}. We can therefore
normalize the unaberrated PSF with the (observed) Strehl ratio, which
is given by the maximum of $p_{\rm obs}$. An equivalent estimate of
the correction factor can be obtained from the requirement for energy
conservation: $\sigma_\Phi^2 = \int y^2+v^2$. This correction to the
equation given by Gonsalves\cite{gonsalves2001} significantly extends
the amplitude of the aberrations over which the algorithm can be
applied. Indeed, an rms aberration of about 1.5 radians can be
tolerated\cite{korkiakoski2012spie}.

\subsection{Odd part of wavefront error}

Once more following Gonsalves\cite{gonsalves2001}, $y$ is easy to
calculate from the odd part of the PSF, Eq.(\ref{eq:po}) via
\begin{equation}
y=\frac{p_o}{2a}\approx\frac{1}{2}\frac{p_oa}{a+\epsilon}\;,
\end{equation}
where $\epsilon>0$ avoids noise amplifications in places where $a$ is
very small. The odd part of the wavefront
error, $\Phi_o(u,v)$ multiplied with the aperture function $A(u,v)$ is
then obtained from the imaginary part of the inverse Fourier transform
of $y$,
\begin{equation}
A\Phi_o = \Im{Y}\;.
\end{equation}

\subsection{Even part of wavefront error}
The magnitude of $v$, the Fourier transform of the even part of
the phase function multiplied with the aperture function can be
easily determined from Eq.(\ref{eq:pesc}), 
\begin{equation}
v=\sqrt{\left|p_e-(1-\sigma_\Phi^2)a^2-y^2\right|}\;.
\end{equation}
We use the absolute value here as this is only approximately correct,
and therefore the argument of the square root has, in practice, also
negative values. Preliminary experiments have shown this to be better
than to set all negative elements to zero. The sign of $v$ cannot be
determined from a single PSF measurement. A convenient way to
determine the sign of $v$ is through phase
diversity\cite{gonsalves1982,gonsalves2001}: an additional, known aberration is
added, very often a defocus term, to figure out the sign of $v$. This
is the topic of the next section. Once a sign has been determined, an
inverse Fourier transform of $v$ leads to the even part of the
wavefront error multiplied with the aperture function,
\begin{equation}
A\Phi_e = \Re{V}\;.
\end{equation}

\begin{figure}
\includegraphics[width=\textwidth]{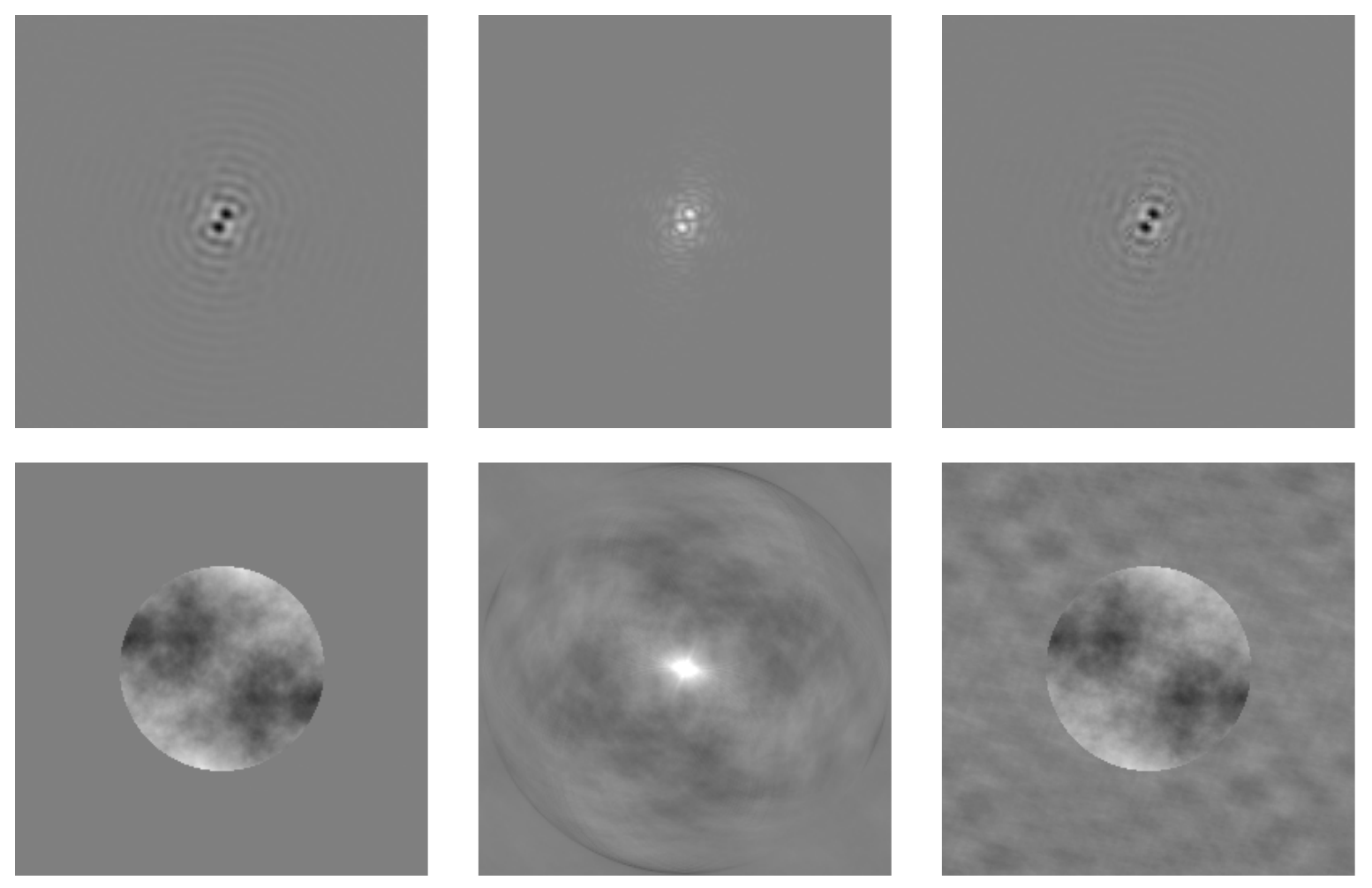}
\caption{The (signed) square root of the even part of the PSF
  due to the even wavefront error, $v$, (top row), along with its Fourier
  transform, $V$, (bottom row), which is an estimate of the even part of the
  wavefront error, $\Phi_e$. The three columns differ in terms of the
  accuracy of the sign of $v$: the correct sign (left column), all
  positive signs (middle column), and 20\% wrong signs (right
  column). Note that sign erros lead to Fourier amplitudes outside of
  the aperture and a corresponding reduction of the amplitudes inside
  the aperture.
\label{fig:sign}}
\end{figure}

Note that the inverse Fourier transform of $v$ should be limited to
the support of the aperture $A$. Any error in the choice of the sign
may lead to Fourier components outside of the aperture, $A$ (see
Fig.~\ref{fig:sign}). Therefore, the sign should be chosen so that the
inverse Fourier transform is limited to the aperture.

\section{Sequential Phase Diversity}
\label{sec:spd}

\subsection{Phase Diversity and Adaptive Optics}

Phase diversity\cite{gonsalves1982, paxman1988, kendrick1994} is a
convenient way to determine the phase aberrations of
images. Typically, two images are recorded simultaneously, often an
in-focus and an out-of-focus image. Another approach, suggested by
Gonsalves\cite{gonsalves2002, gonsalves2010} consists in using a
deformable mirror in an adaptive optics system to sequentially
introduce an additional phase aberration. It is indeed easy to see how
an adaptive optics system in normal operations can be used as a
sequential phase diversity wavefront sensor. Imagine that at time
$t_k$ we observe a partially corrected image that has an (unknown)
phase aberration $\Phi_k$. If we can assume that the deformable mirror
can change its shape faster than the typical changes in the incoming
wavefront that is being corrected, then the deformable mirror of the
AO system in the previous update cycle had introduced an additional
phase aberration of $-\Delta\Phi_{k-1}$ where $\Delta\Phi_{k-1}$ is
the correction that was made by the deformable mirror in the last iteration.

To use phase-diversity as a wavefront sensor in an adaptive optics
system, one therefore has to know the previous and the current
focal-plane images along with the phase change introduced by the
deformable mirror during the previous update cycle.

\subsection{Choice of diversity}

In the weak-aberration regime developed above, phase diversity only
has to help us find the sign of $v$. In most cases, one chooses a
quadratic diversity as it is easily implemented by defocusing the
imaging camera. When using a deformable mirror, much more elaborate
choices of diversity can be chosen. Indeed, our goal is to find a
diversity for every update cycle that minimizes the computational
effort of determining the sign of $v$.

If one uses a small diversity, it is easy to
show\cite{gonsalves2001} that the even part of the PSF is
\begin{equation}
p_e^\prime = (1-\sigma_\Phi^{\prime 2})a^2 + y^2 + (v+z)^2\;,
\label{eq:div1}
\end{equation}
where $z=a*\delta\phi_e$, where $\delta\phi_e$ is the even part of
$\delta\phi$, which in turn is the Fourier transform of
$\Delta\Phi$. If we know the introduced diversity, we can easily
calculate $z$. Note that the introduced diversity does not have to be
even, as is almost always assumed in phase-diversity work, since odd
and even wavefront errors are nicely separated in the odd and even
parts of the PSF. It is therefore sufficient to look at the even part
of the introduced phase diversity.

A simple approach to minimize the computational effort is the
following: start by choosing an arbitrary sign for $v$ and call it
$v_{k-1}$.  In the next update cycle of the AO system, we make the
assumption that $z=-v_{k-1}$ (note that this is not true since $z$ is
related to the introduced diversity, not what we think that we
introduced. Therefore, if the choice of sign was correct,
$(v_k-v_{k-1})^2=0$, and if the sign was wrong,
$(v_k-v_{k-1})^2\approx 4v_{k-1}^2$. The prescription is then to
monitor $p_e$ at every point in the focal plane. If the magnitude of
$(p_{e,k}-y_k^2)(x,y)>(p_{e,k-1}-y_{k-1}^2)(x,y)$ then one flips the
sign at location $(x,y)$, i.e. $\mathrm{sign}\; v_k=-\mathrm{sign}\;
v_{k-1}$. The problem with this simple approach is two-fold: 1) the
algorithm in no way takes into account that the inverse Fourier
transform of $v$ should have a support limited by $A$; 2) once the
correct sign choice has been made, the wavefront error will not be
zero because of all the approximations one has made, and in particular
the Fourier issue mentioned under 1). The choice in sign has again to
be a random choice, which reduces the convergence speed by about a
factor of 2 from what we would like to achieve.

A significantly better approach is modeled after the one introduced by
Gonsalves\cite{gonsalves2001} where we take into account that we also
update the odd wavefront, and not just the even wavefront. The sign of
$v_k$ in iteration $k$ is then determined by
\begin{equation}
\mathrm{sign }v_k = \mathrm{sign}(v_{k-1}^2-v_{k}^2-z^2)/(2z)\;.
\label{eq:vsign}
\end{equation}
This approach has two significant advantages, possibly not realized by
Gonsalves\cite{gonsalves2001}: 1) it deals with the Fourier support
issue as $v_{k-1}$ does not necessarily obey the Fourier support
contraint, but $z$ does, and 2) even if the correct sign choice was
made for $v_{k-1}$, there will be a residual error, and the above
equation also tries to estimate the best choice in this iteration
based on taking into account the difference between $v_{k-1}$ and $z$.

The problem is equivalent to being able to finding the minimum of
$y=x^2$ in the case that we can change $x$, but only measure $y$. We
measure an initial value of $x_1^2$. Without any additional
information, we must randomly choose the sign of the step that we are
going to take to minimize $y$. Without loss of generality, we choose
$\delta x_1 = -\sqrt{x_1^2}$, i.e.\ we assume that we are on the right
side of $x=0$. We then measure a value of $x_2^2$. Obviously, if
$x_2^2>x_1^2$, we chose the wrong sign. If we chose the correct sign,
then the next step is $\delta x_2=\pm\sqrt{x_2^2}$. But we would also
like to know on which side of $x=0$ we are after having taken the step
$\delta x_1$ so that we do not have to again guess the sign in the
next iteration. The correct sign for $\delta x_2$ can be estimated
form the following relation:
\begin{equation}
x_1^2 = (x_2+\delta x_1)^2 = x_2^2+2x_2\delta x_1+\delta x_1^2\;.
\end{equation}
Therefore the sign of $x_2$ is given by the following equation
\begin{equation}
x_2 = (x_1^2-x_2^2-\delta x_1^2)/(2\delta x_1)\;.
\end{equation}
If $x_2$ is positive, $\delta x_2$ must be negative, and vice versa.

Even if the commanded value of $\delta x_1$ and the actually achieved
step in $x$ are not the same, the above equation is still correct when
replacing $\delta x_1$ with the actual step $z$, which is then
equivalent to Eq.(\ref{eq:vsign}). This is particularly important in
our case as we will always have errors in the choice of sign for some
parts of $v$. Because of these errors, the deformable mirror will not
achieve the correct shape, and the actual change will lead to a step
size $z$ that is different from what we commanded initially.

\section{Reducing computational complexity}
\label{sec:complex}

An obvious way to minimize the computing requirements consists in
reducing the number of Fourier transforms. Due to the symmetry
properties of Fourier transforms, the inverse
Fourier transforms of $y$ and $v$ can easily be combined into a single,
complex Fourier transform according to 
\begin{equation}
A(u,v)^\prime\Phi^\prime(u,v) =\Re{(V+Y)}+\Im{(V+Y)}\;.
\end{equation}

The calculation of $z$ also requires two Fourier transforms to
implement the pupil plane support restriction. Realizing that this 1)
low-pass filtering in the Fourier domain is nothing more than a
two-dimensional filter in the focal plane (a sinc filter in the case
of a top-hat aperture function $A(u,v)$) and 2) that this filter in the
focal plane is strongly concentrated around the origin, we can perform
this filtering with a brute-force convolution with a very limited
kernel of about 5 by 5 pixels. This will add about 25 multiply-add
calculations per actuator, which is negligible compared to the $NlogN$
complexity of a fast Fourier transform.

Since the limited support of $V$ in the Fourier domain is
nothing more than a low-pass filter applied to $v$, one can compare
the estimate of $v$ one has obtained and compare it to the convolution
of $v$ with the Fourier transform of $A$, which is $v*a$. If $v$ and
$v*a$ have the same sign, then the chosen sign for $v$ is correct,
otherwise the opposite sign is a better choice. This {\em
  post-processing} step can again be implemented with a direct
convolution with a very small kernel.

\begin{figure}
\includegraphics[width=\textwidth]{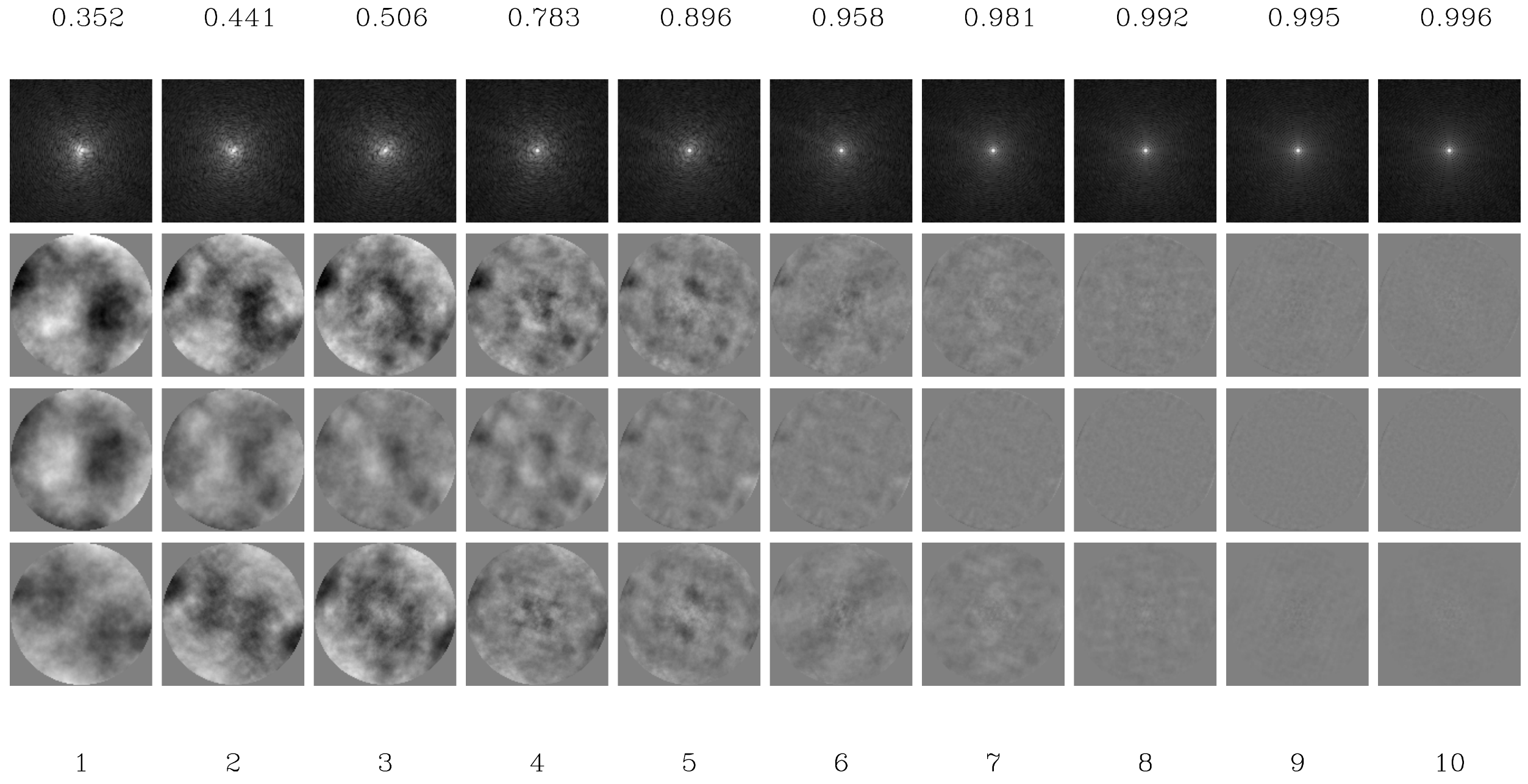}
\caption{Example of the convergence of our algorithm in terms of the
  PSF (top), total wavefront error (second from top), odd wavefront
  error (second from bottom) and even wavefront error (bottom) along
  with the iteration number at the very bottom and the Strehl ratio at the top.
  \label{fig:iter}
}
\end{figure}

Figure~\ref{fig:iter} shows 10 iterations of our algorithm with the
direct convolution approximation for $z$ and the post-processing of
the sign choice as described above for a static, Kolmogorov-spectrum
aberration with a starting Strehl ratio of 0.35.

\section{Advantages}
\label{sec:adv}

The obvious advantage of our approach is that it only requires a
single, complex Fourier transform and a number of operations
proportional to the number of actuators. In contrast to other Fourier
reconstructors\cite{poyneer2002}, our approach needs no particular
treatment of the edges as we are directly sensing the phase, not its
spatial derivative. Computationally, it is therefore very fast and can
easily cope with systems that have 10,000 to 100,000 actuators. 

Some phase retrieval algorithms have a tendency to stall. This is not
the case here as any increase in Strehl ratio will also mean that the
weak-aberration assumption is more accurate, and therefore the
algorithm will perform better.

Our algorithm is also ideally suited for sensing a wavefront after an
apodizing pupil coronagraph. The odd term and the second term in the
even part of the PSF are both products with $a$, the complex field
corresponding to the perfect PSF. As the main task of a coronagraph is
to reduce the magnitude of $a$, these two terms will also decrease
correspondingly. The estimation of the even part of the wavefront
therefore becomes much easier since the second term contributes much
less.
 
We can also limit the radial extension of the PSFs that we work with, and can
thereby limit the number of Fourier modes that are reconstructed. Even
an obscuration from a coronagraph image mask can be taken into account
if the corresponding low-order aberrations are well corrected.

Finally, our approach easily implements any Fourier-based technique to
make predictive corrections, e.g.\ shifting the entire wavefront in
one direction, or take into account the influence function of
actuators as long as it can be described by a convolution.

\section{Limitations}
\label{sec:lim}

Our wavefront sensing approach has been developed for monochromatic
light. Broadband light leads to a radial smearing of the wavefront
information as the speckle pattern scales with wavelength
$\lambda$. One way to overcome this problem is the use of an extremely
chromatic optical system, which has a wavelength-dependent
magnification such that the scaling with $\lambda$ due to diffraction
is exactly compensated. Such optical systems have been described in
the literature\cite{wyne1979, roddier1980}. The broadband wavefront
sensing will therefore require its own optical system, and we cannot
directly use the science focal-plane for the wavefront sensing.

The algorithm is currently limited to a single point source, which is
often the case for exoplanet imaging, and can easily be implemented in
laboratory applications. An application to extended objects is
feasible, but it is not clear at this time that all wavefront modes
can indeed be measured.

Another requirement is that the aperture has an even symmetry. It
therefore also works with typical amplitude-apodization pupil masks,
but will not work with typical pupil phase-mask coronagraphs, which
are even in their real part and odd in their imaginary part. It might
be possible to modify our approach to phase-mask coronagraphs, which
will be part of our future efforts.

We have not optimized our approach in any ways regarding the
different free parameters or the gains in the feedback loop. Such
optimizations will be the topic of future investigations. And,
finally, we have not yet considered the influence of amplitude
aberrations and deviations between the expected and the true, perfect,
PSF on our algorithm. The successful laboratory experiments performed by our
team\cite{korkiakoski2012spie}, however, give us hope that these
potential errors may be overcome.

\section{Applications}
\label{sec:apps}

The original idea for this development was driven by the need for an
efficient wavefront sensing and extreme adaptive optics control
algorithm for high-contrast imaging that scales more gracefully than
the classical number-of-actuators-squared ($N^2$) law. Such an
algorithm is indeed essential for future exoplanet imaging instruments
for the next generation of extremely large telescopes. Apart from
this originally intended application, our approach is useful for many
other applications such as non-common path aberration measurements and
slow shape corrections for future high-contrast imaging space missions. 

Indeed, our approach is ideally suited to measure small, non-common path
aberrations in AO-assisted instrument. To determine them, one uses a
monochromatic point source in a focus in front of the AO system. Such
a light source is often already present to calibrate the adaptive
optics itself. Our approach now provides for an iterative wavefront
sensing in the focal plane using the deformable mirror to introduce
the required diversity. Results from an actual test of this approach
can be found in these proceedings\cite{korkiakoski2012spie}.


\bibliography{KellerSPIE2012}   
\bibliographystyle{spiebib}   

\end{document}